\documentclass[12pt]{iopart}

\usepackage{iopams}  
\usepackage{graphicx}	
\usepackage{bm} 
\usepackage{color}
\usepackage{setspace}
\usepackage{placeins}

\begin{document}
\title{Static and dynamic phases of a Tonks-Girardeau gas in an optical lattice}
\date{\today}
\author{Mathias Mikkelsen$^1$, Thom{\'a}s Fogarty$^1$ and Thomas Busch$^1$}

\address{$^1$Quantum Systems Unit, OIST Graduate University, Onna, Okinawa 904-0495, Japan}
\ead{\mailto{mathias.mikkelsen@oist.jp}}
 
\begin{abstract} 
We investigate the properties of a Tonks-Girardeau gas in the presence of a one-dimensional lattice potential. Such a system is known to exhibit a pinning transition when the lattice is commensurate with the particle density, leading to the formation of an insulating state even at infinitesimally small lattice depths. Here we examine the properties of the gas at all lattices depths and, in addition to the static properties, also consider the non-adiabatic dynamics induced by the sudden motion of the lattice potential with a constant speed. Our work provides a continuum counterpart to the work done in discrete lattice models.
\end{abstract}

\maketitle

\section{Introduction}
\label{sec:intro}

The versatility of modern ultracold atomic experiments has in recent years allowed for the realisation of exotic matter wave phases borne from the competition between different physical forces acting on many-body systems \cite{Sachdev2008,Sachdev,Vojta}. The paradigmatic example of this has been the observation of the superfluid-to-Mott-insulator phase transition in ultracold gases confined in three-dimensional optical lattice potentials, which stems from the interplay between the tunnelling dynamics and the onsite interactions  \cite{Greiner,Jordens:08}. In one dimensional lattices and for strong repulsive interactions it has been shown that the phases of a system are entirely controlled by the particle filling statistics \cite{Zwerger2003,Haller2010,Lelas2012,Cartarius2015,Hoogerland2016}. In this case, a commensurate particle density leads to an integer filling of each lattice site and thus realises an insulating \textit{pinned} phase where dynamics is suppressed due to vanishing long range coherence. For incommensurate particle densities each lattice site has a non-integer filling which has the effect of creating delocalised modes which can be considered as defects \cite{Astrakharchik2016,Astrakharchik2017}. These defects promote dynamics as they preserve a degree of coherence and allow the system to attain superfluid-like properties. Quantum phase transitions in these systems can be probed by switching the control parameter across a critical point \cite{Silva2008,Karrasch2013,Campbell2016,Heyl2017,Fogarty2017,Sengstock2017}. Additionally, the transport properties of such systems can be linked to classical models of friction which describe stick-slip motion between two surfaces \cite{Peyrard1982,Braun2013,Mandelli2013,Fogarty2015,Fogarty2016}, an effect which has been recently observed \cite{Vladan2016,Vladan2017,Kont2018}. In fact, precise control of the competing length scales between the two surfaces can be used to effectively control the amount of friction present\cite{Vladan2015} and can give insights into designing frictionless dynamical processes for quantum systems. 

In this work we focus on describing the static and dynamical phases of an ultracold gas in an optical lattice potential with periodic boundary conditions \cite{Cazalilla2003,Brouzos2010,Lazarides2012}. We focus on the Tonks-Girardeau (TG) limit in one dimension \cite{Paredes2004,Kinoshita2005}, where the appearance of the pinned phase can be observed for particle numbers commensurate with the underlying lattice, such that the system becomes insulating even for very shallow lattice depths.  Suddenly setting the lattice potential into motion with a constant rotation speed allows us to probe 
the dynamical response of the TG gas to a quench in the external potential and to study the relationship between between friction and commensurability. The observed behaviour is similar to a type of quantum sprocket, where the rotating lattice will impart maximum angular momentum to the gas if the system is in the insulating phase, while in the superfluid phase only reduced momentum transfer is observed due to tunneling of the particles between lattice sites. Our work is related to recent investigations of driven cold atom systems in the continuum \cite{Citro2009, Hallwood2010,Schenke2011,Schenke2012,Schloss2016, Minguzzi2017}, and in the Bose-Hubbard model \cite{Pinheiro2013,Arwas2017,Minguzzi2017b,Minguzzi2018}, which is naturally connected to the continuous model in the limit of tight trapping. Another recent related work has investigated the localization properties of a TG gas in a continuum version of the Aubry-Andre model (using a bi-chromatic lattice) \cite{Settino2017}. In this manuscript we explore the phase diagram from shallow to deep lattices for commensurate and incommensurate fillings and show how different phases can be characterized through the coherence and by direct observation of the momentum distribution. We also show that the dynamical response of the system to the driven lattice exhibits stick-slip motion and tunneling due to defect induced superfluidity.

The mansucript is organised as follows: In Section \ref{sec:model} we describe the model we consider and in Section \ref{sec:statics} we discuss how to characterise the different phases in the system through calculations of the coherence, the momentum distribution and the spatial auto-correlation function. In Section \ref{sec:driven} we discuss both the time averaged and the instantaneous dynamics of the driven system and in Section \ref{sec:conclusions} we conclude. Finally, in the Appendix we provide an in depth explanation of the dynamics in the different phases.

\section{Basic Model}
\label{sec:model}
We consider a one-dimensional gas of $N$ neutral bosons of mass $m$ which are subject to an external potential $V_\mathrm{ext}(x_n)$. The Hamiltonian is given by
\begin{equation}
  H= \sum^N_{n=1} \left[ -\frac{\hbar^2}{2m}\frac{\partial^2}{\partial x_n^2}+V_\mathrm{ext}(x_n) \right]+ g \sum_{i<j}\delta(|x_i-x_j|),
\label{eq:basicmodel}
\end{equation}
where the two-body interaction between the particles is described by a zero-range delta-function potential of strength $g$ \cite{Olshanii:98}. While exact solutions for this Hamiltonian exist in the absence of an external potential \cite{Sutherland:04,Franchini:17}, only limiting cases can be solved in other situations. 

One regime in which the Hamiltonian \eref{eq:basicmodel}  can be solved exactly is the strongly interacting Tonks-Girardeau limit. It corresponds to setting $g=\infty$, which allows one to obtain an exact solution for arbitrary $N$ by utilizing the Bose-Fermi mapping theorem \cite{Girardeau1960}. The basic idea behind this theorem is to replace the interaction term by a constraint on the many-body wavefunction of the form 
\begin{equation}
   \Psi(x_1,...x_N) =0 \qquad  \mathrm{if} \qquad |x_i-x_j|=0,
\end{equation}
which is equivalent to enforcing the Pauli exclusion principle in position space. The strongly repulsive bosonic system can therefore be mapped onto a gas of non-interacting, spinless fermions for which the many-body wave-function can be written as a sum of single particle product states using the Slater determinant
\begin{equation}
  \Psi_F(x_1,...,x_N)=\frac{1}{\sqrt{N!}} \mathrm{det}^{(N,N)}_{(n,i)=(1,1)}\psi_n(x_i).
\end{equation}
Here the $\psi_n(x)$ are the eigenfunctions of the corresponding single-particle Hamiltonian
\begin{equation}
H_\mathrm{SP}= -\frac{\hbar^2}{2m}\frac{\partial^2}{\partial x^2}+V_{\mathrm{ext}}(x),
\label{eq:SP}
\end{equation}
for which exact values can easily be obtained numerically using finite-difference diagonalization. The bosonic many-body wavefunction can then be obtained from the fermionic one by symmetrising with an anti-symmetric unit function of the form $A(x_1,...,x_N)= \Pi_{1\leq i <  j \leq N} \mathrm{sgn}(x_i-x_j)$. This results in  
\begin{equation}
\Psi_B(x_1,\dots,x_N)=A(x_1,\dots,x_N)\Psi_F(x_1,\dots,x_N).
\end{equation}

We chose the external potential to be a periodic lattice of the form $V_\mathrm{ext}(x_n)=V_0 \cos^2(k_l x_n)$, where $V_0$ is the lattice depth and $k_l$ the wavevector. We consider this trapping potential to possess periodic boundary conditions such that it realizes a 1D ring trap of length $L$ with regular intensity modulations and  $\Psi(...x_n+L...)=\Psi(...x_n...)$ for any $n$. This allows us to write the lattice wavevector in terms of the number of wells in the lattice potential $M$ as $k_l =  \pi M/L$ and provides an energy scale for the system in terms of the recoil energy $E_r=\frac{\hbar^2 \pi^2 M^2}{2 m L^2}$.
For consistency we will use systems with $M=50$ lattice sites for all calculations shown in this manuscript.

It is important to note that the requirement to have periodic boundary conditions for the bosonic TG gas means that the fermionic many-body wavefunction must have periodic boundary conditions (PBC) for odd $N$ and anti-periodic boundary conditions (A-PBC) for even $N$, as the anti-symmetric unit function leads to $\Psi_F(...x_n+L...)=(-1)^{N-1}\Psi_F(...x_n...)$. Since the single-particle eigenfunctions, $\psi_n(x)$, form an orthogonal basis, the Slater determinant obeys these boundary conditions if 
\begin{equation}
  \psi_n(L)=(-1)^{N-1} \psi_n(0).
  \label{eq:periodicBC}
\end{equation}

\begin{figure}[tb]
\centering
\includegraphics[width=0.7\linewidth]{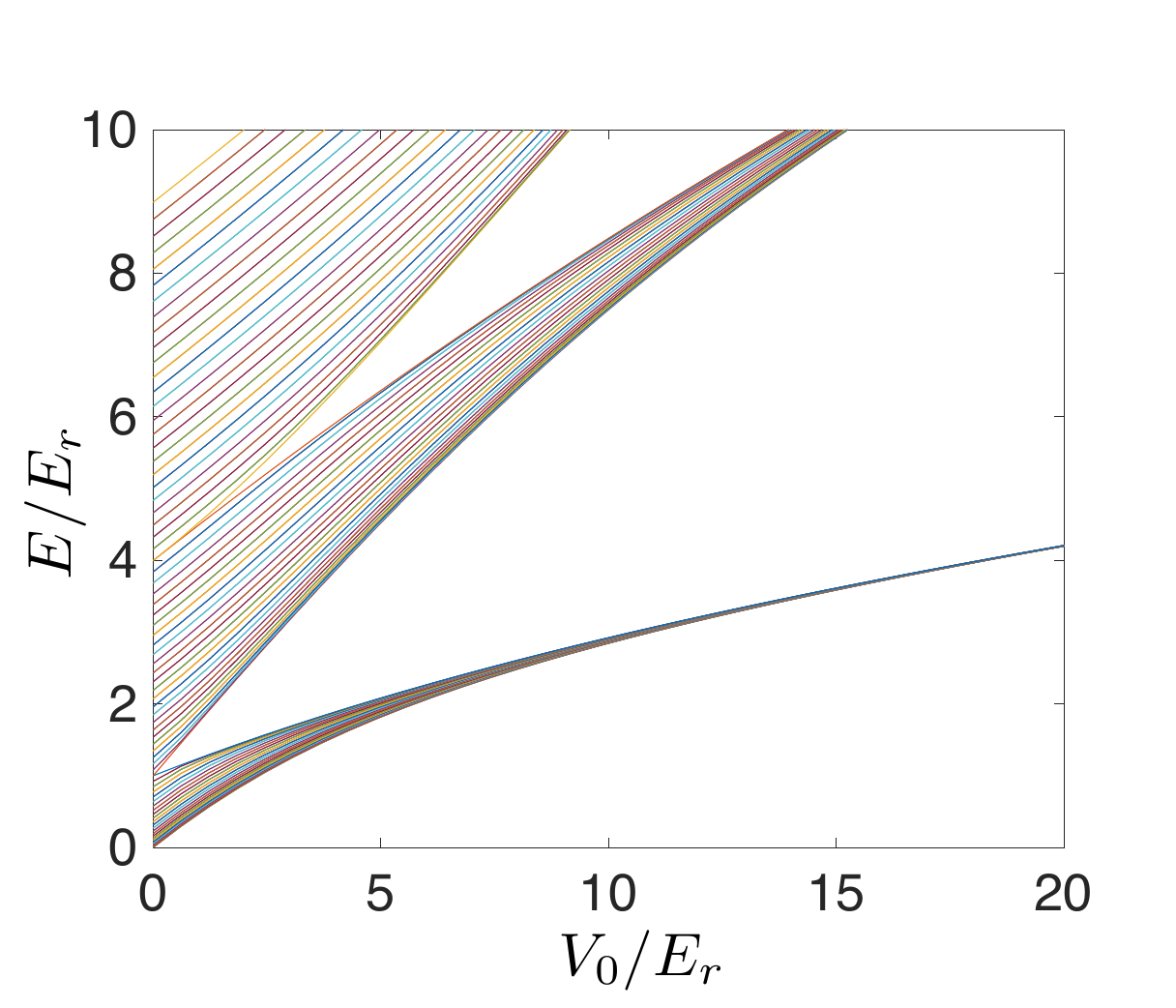}
\caption{Lowest lying 150 single-particle eigenenergies as a function of lattice depth for $M=50$.} 
 \label{fig:SPspectrum} 
\end{figure}

In free space the single-particle solutions are given by  $\psi_n(x) = \frac{1}{\sqrt{L}} e^{i k_n x}$, with $E_n=\frac{\hbar^2 k_n^2}{2m}$.  For PBC (corresponding to an odd number of particles) we therefore have $k_n = 2 n \pi /L$ with $n=\lbrace 0,\pm 1,\pm 2,\pm 3 \dots\rbrace$. In this case the ground-state is non-degenerate, while all excited states $E_n,E_{-n}$ are two-fold degenerate. For A-PBC (corresponding to an even number of particles) all states are two-fold degenerate as $n=\lbrace \pm 1,\pm 2,\pm 3 \dots\rbrace$. The effect of the lattice is to introduce energy gaps into the single particle spectrum (see Fig.~\ref{fig:SPspectrum}), and while the free space degeneracy structure remains for a large range of lattice depths, this gap can break the degeneracy between the two states closest to the gap. For example, for an even number of sites this will be the case in the PBC spectrum, but not in the A-PBC spectrum. A more detailed discussion of this odd-even effect is given in Section \ref{sec:driven}, where it becomes physically relevant. In the deep lattice limit the eigenstates become M-fold degenerate as individual lattice sites effectively decouple. To avoid numerical issues and have a clearly-defined ground-state Slater determinant, we avoid this limit and stay within a range of lattice depths (0 to 20 $E_r$) for which the first $M$ states are not yet completely degenerate.

\section{Statics of the TG gas}
\label{sec:statics}
In the following we will first discuss the ground state properties of the TG gas in the presence of a lattice.

\subsection{The pinned and the superfluid phase}
The Hamiltonian \eref{eq:basicmodel} is known to possesses two distinct ground-state phases in the TG limit, which appear as a function of the ratio of number of bosons to lattice sites
\begin{equation}
F=\frac{N}{M}\;.
\end{equation}
For incommensurate fillings ($F \neq \mathcal{N}$, with $\mathcal{N}$ a positive integer) in shallow lattices the system has superfluid-like characteristics with long-range coherence and good conductivity due to the delocalisation of the wavefunction over many lattice sites. However for commensurate filling ($F= \mathcal{N}$), the bosons becomes localised at individual lattice sites and the total system becomes \textit{pinned} to the lattice. This pinned phase has no coherence, behaves as an insulator and is the hard-core continuum analogue of the Mott-insulator phase in the Bose-Hubbard model (BHM). In the continuum TG model the pinning happens at infinitesimally small lattice depths for $F=1$. It was first theoretically proposed by B\"uchler \textit{et al.} \cite{Zwerger2003} and has been experimentally observed in numerous cold atom experiments \cite{Haller2010,Hoogerland2016}.

The degree of coherence is a good property to characterize the two different phases of the system. It can be obtained from the off-diagonal elements of the reduced single-particle density matrix (RSPDM)
\begin{equation}
\rho_1(x,y) = \langle \Psi_B | \hat{\psi}^{\dagger}(x) \hat{\psi}(y) | \Psi_B \rangle,
\end{equation}
which describes the spatial auto-correlation of a single particle, giving the probability that a particle is at $y$ immediately after it has been measured at $x$. For classical particles and for non-interacting fermions, the RSPDM will always be diagonal, however for a Bose gas the off-diagonal elements play a prominent role. Efficient algorithms to numerically calculate the RSPDM in the TG regime exist \cite{Pezer2007}, which are tractable even for large numbers of particles and which we have used in our calculations.

Diagonalising the RSPDM leads to a set of eigenstates $\phi_i(x)$, known as the natural orbitals, with the corresponding eigenvalues $\lambda_i$ giving the respective occupation numbers
\begin{equation}
   \rho_1(x,y) = \sum_i \lambda_i \phi_i^*(x) \phi_i(y),
\label{eq:naturalorbitals}
\end{equation}
where $\sum \lambda_i=N$. For a non-interacting Bose gas at zero temperature only the lowest lying orbital is occupied, $\lambda_0=N$, which corresponds to a completely coherent and superfluid system. 
Conversely, a ground state wavefunction with $\lambda_0=1$ corresponds to a completely incoherent gas where the $N$ lowest orbitals are equally occupied. The occupation number of the lowest lying eigenstate of the RSPDM can therefore be used as a measure of the coherence of the system. 
In fact, for a TG gas in free space it is known that the lowest lying orbital has an occupation proportional to $\sqrt{N}$, while the occupation of orbitals with $i\gg N$ tends to zero \cite{Lenard1964,Forrester2003}. Clearly such a state is not strictly superfluid, however it is also not completely incoherent, as $\lambda_0$ is of the order $\sqrt{N}$. It therefore does possess some superfluid properties. For simplicity, we will refer to phases displaying finite coherence as superfluid in the rest of this manuscript. In the pinned phase, however, each particle is highly localised which destroys the coherence between individual wells once tunneling is sufficiently suppressed. The system is therefore reduced to $N$ non-interacting particles with coherence $\lambda_i=1$ for the $N$ lowest lying orbitals.  

Even though the coherence can be used to identify different regimes of the system, it is not easy to measure it in a direct way. However, it is closely linked to the momentum distribution, which can be readily observed through time-of-flight experiments and calculated from the Fourier transform of  $\rho_1(x,y)$ as
\begin{equation}
 n(k) = \int dx \, dy \rho_1(x,y) e^{-i k(x-y)}.
\end{equation}
This can be recast in terms of the Fourier transform of the eigenstates of the RSPDM, $\widetilde{\phi_i}(k)$, using the same occupation numbers 
\begin{equation}
  n(k) = \sum_{i} \lambda_i \widetilde{\phi_i}^*(k) \widetilde{\phi_i}(k).
\label{eq:mom} 
\end{equation}
It is therefore clear that the momentum distribution depends directly on the distinct characteristics of the coherence and long-range order of the system. 

\subsection{Coherence and momentum distribution}

\begin{figure}[tb]
\centering
\includegraphics[width=\linewidth]{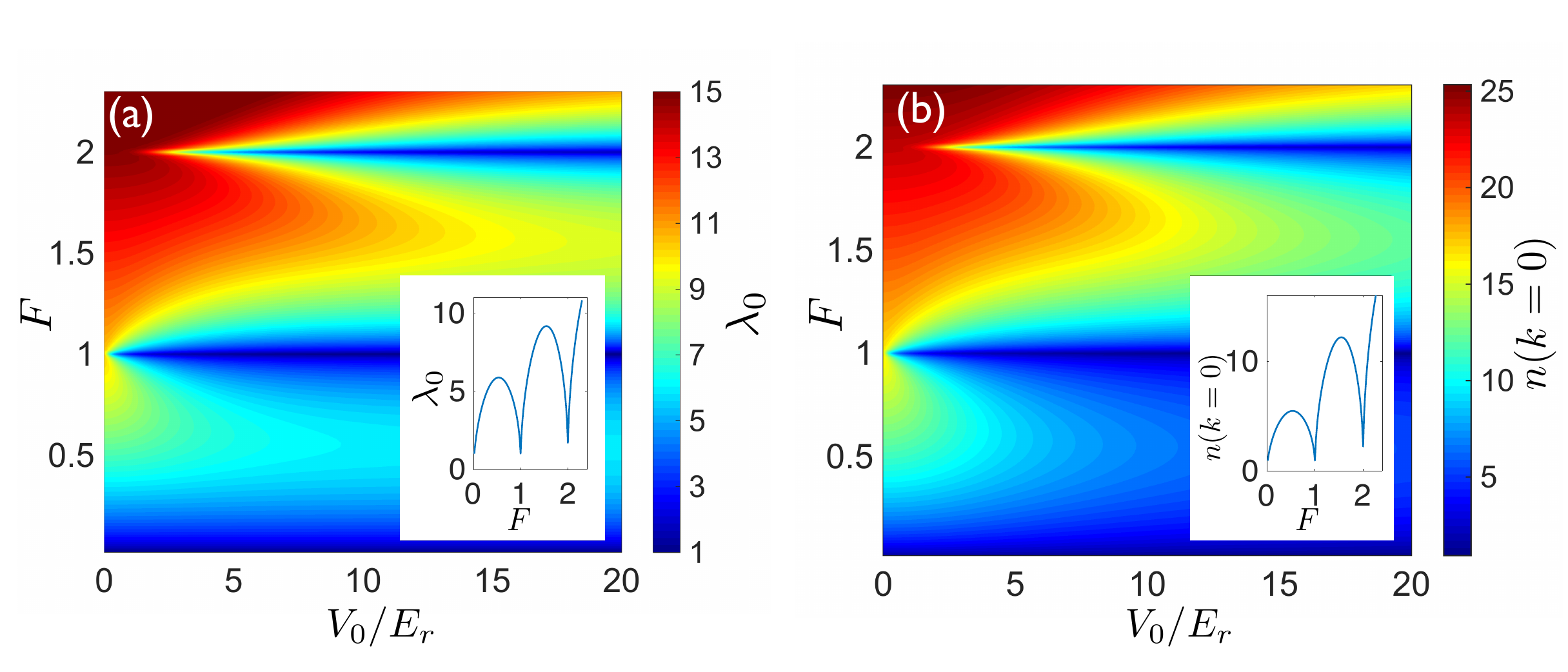}
\caption{(a) Coherence $\lambda_0$ and (b)  height of the central peak of the momentum distribution $n(k=0)$ as a function of the filling ratio $F$ and the lattice depth $V_0$ (in units of $E_r$). The number of lattice wells is fixed at $M=50$. The insets shows the coherence and the zero momentum peak as a function of $F$ for $V_0=20 E_r$.} 
 \label{fig:staticphasediagram} 
\end{figure}

The lowest eigenvalue of the RSPDM, $\lambda_0$, and the peak-value of the momentum distribution, $n(k=0)$, are shown in Fig.~\ref{fig:staticphasediagram} as a function of the filling ratio and the lattice depth. One can immediately see that the coherence decreases quickly for unit filling, $F=1$, eventually reaching the value of $\lambda_0=1$ corresponding to the pinned phase. 
For double filling, $F=2$,  a finite lattice depth is required before this transition can happen, as the lattice needs to be deep enough to overcome the repulsive two-body interactions at each site. For incommensurate fillings a slow decrease of the coherence with increasing lattice depth can be seen. This  behavior is qualitatively mirrored in the magnitude of the momentum distribution at $k=0$ and in the deep lattice limit ($V_0=20 E_r$), both quantities display an oscillating dependence on the filling ratio $F$, as can be seen in the insets of Fig.~\ref{fig:staticphasediagram}. This is similar to the behaviour known for the superfluid fraction in the hard-core BHM \cite{Krutitsky2016}.

\begin{figure}[tb]
\includegraphics[width=\linewidth]{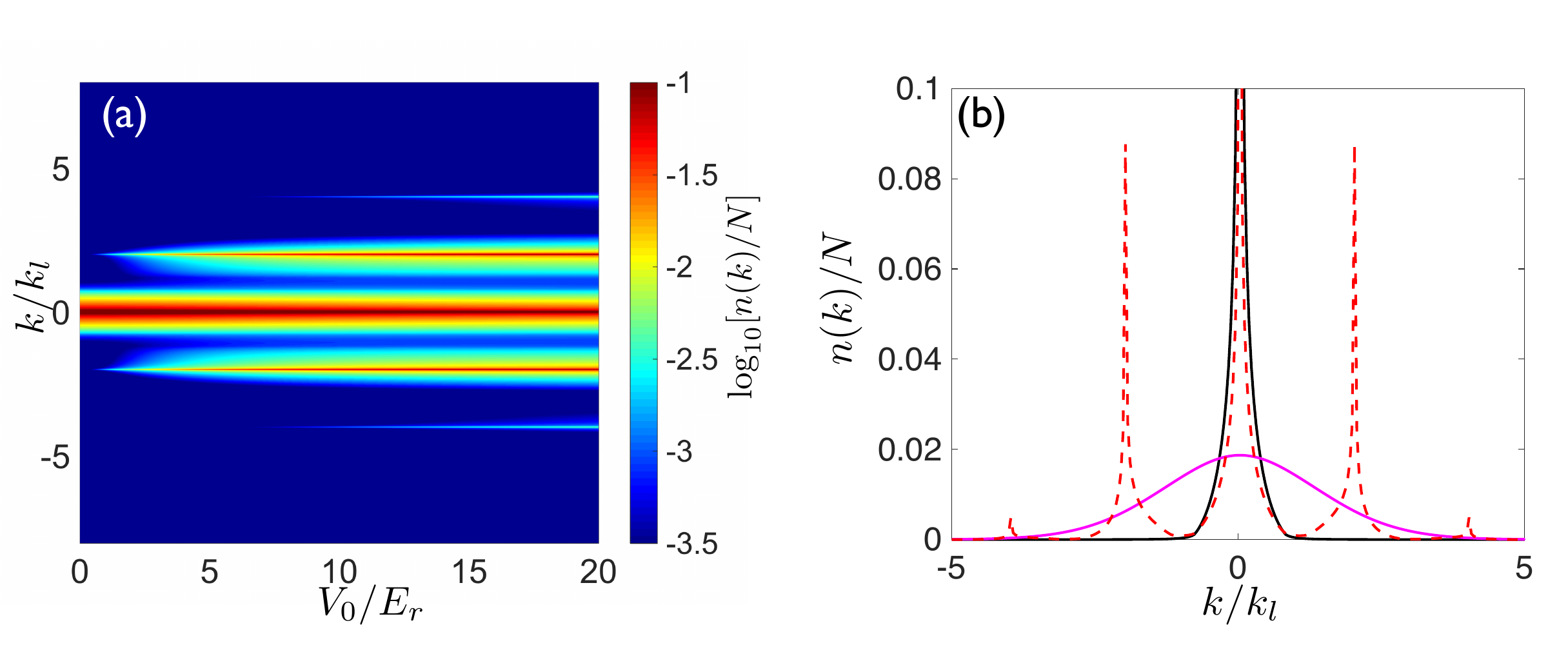} 
\caption{(a) Normalised momentum distribution for an incommensurate state with $F=\frac{2}{5}$ as a function of lattice depth. (b) Normalised momentum distribution in a deep lattice ($V_0 = 20 E_r$) for the incommensurate ($F=\frac{2}{5}$, red dashed line) and the commensurate ($F=1$, full magenta line) case. For comparison the black line corresponds to the case where no lattice is present.} 
 \label{fig:momentumdistributionasfunctionofdepth} 
\end{figure}

The distinct characteristics of the pinned phase and the superfluid phase also manifest themselves in the full momentum distribution. In order to characterize the incommensurate phase we show the momentum distribution for $F=\frac{2}{5}$ as a function of the lattice depth in Fig.~\ref{fig:momentumdistributionasfunctionofdepth}(a) and with a comparison to the commensurate case in Fig.~\ref{fig:momentumdistributionasfunctionofdepth}(b).
In free space ($V_0=0$) the TG gas is delocalised in position space and therefore localises around a single, central peak in momentum space (black solid line in Fig.~\ref{fig:momentumdistributionasfunctionofdepth}(b)), 
whereas for $V_0=20$ in the pinned phase ($F=1$) the gas is localised in position space and is therefore delocalised in momentum-space (magenta line). In the incommensurate phase ($F=\frac{2}{5}$) 
the momentum distribution can be seen to display additional back-scattering peaks at multiples of $2 k_l$, the number of which and their intensity increases with the lattice depth. The gas in this phase therefore represents a momentum delocalised superfluid that exhibits some of the distinct characteristics of both the superfluid and the pinned phase. The observed momentum distribution is similar to the momentum distributions found in supersolids \cite{Boninsegni2012} and it has recently been suggested that the incommensurate phase of the TG gas in deep lattices is similar to a defect-induced superfluid phase and can be utilized to investigate the Andreev-Lifschitz-Chester mechanism \cite{Cinti2014,Astrakharchik2017}. 

\begin{figure}
\centering
\includegraphics[width=\linewidth]{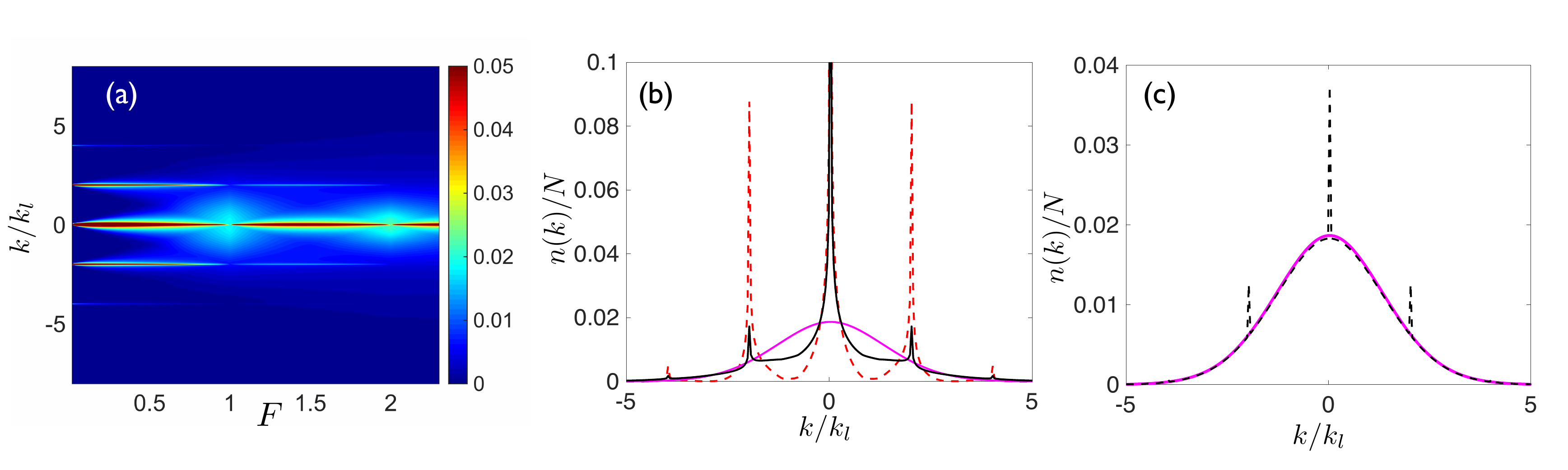} 
\caption{ (a) Normalized momentum distribution as a function of the filling factor. (b) Momentum distribution for filling factors $F=1$ (magenta line), $F=\frac{2}{5}$ (dashed red line) and $F=\frac{7}{5}$ (black line). (c) Momentum distribution for filling factors $F=1$ (magenta line) and $F=\frac{49}{50}$ (dashed black line). All plots in this figure are for a lattice depth of $V_0 = 20 E_r$.} 
 \label{fig:momentumdistributionasfunctionofratio} 
\end{figure}

It is instructive to consider two distinct defect regimes of the incommensurate phase. If the deviation from the commensurate phase at $F=1$ is microscopic, such that $N=M\pm1$,  the phase becomes gapless with a quadratic dispersion relation.  On the other hand, for a macroscopic number of defects, such as $N=M\pm M/2$,  the phase is also gapless, but with a linear dispersion relation \cite{Astrakharchik2017}. These regimes are, however, continuously connected which can be seen in Fig.\ref{fig:momentumdistributionasfunctionofratio}(a), where the momentum distribution as a function of $F$ is plotted for a constant depth $V_0 = 20 E_r$. Far from commensurate values, i.e.~for a macroscopic number of defects, the multiple peaks stemming from the delocalised superfluid are the dominant contribution to the momentum distribution (see Fig.~\ref{fig:momentumdistributionasfunctionofratio}(b)), but as $F\rightarrow 1$ a Gaussian contribution corresponding to the pinned phase momentum distribution becomes more dominant (see Fig.~\ref{fig:momentumdistributionasfunctionofratio}(c)). In fact, for a microscopic numbers of defects, such as for $N=M\pm 1$, the momentum distribution essentially corresponds to that of the delocalised pinned phase, but with additional small peaks on top of it. Beyond unit filling the Gaussian contribution of the first $N=M$ particles remains, while the additional particles lead to a multi-peak structure on top of it. This means that the first $N=M$ particles are effectively filling the lowest energy-band, while the superfluid physics happens in the second band. 

\subsection{Spatial auto-correlations}

The multi-peak structure observed in the momentum distribution of the incommensurate TG gas is a clear indicator that off-diagonal long-range order (ODLRO) and density long-range order (LRO) exist in the gas. To confirm this we calculate the spatial auto-correlations, which we define at a distance $r$ as the average of the RSPDM-values at $(x,y)$, i.e.
\begin{equation}
  \rho_1(r) = \frac{1}{L^2}\int dx dy \rho_1(x,y) \delta(|x-y|-r).
\end{equation} 

\begin{figure}[tb]
\centering
\includegraphics[width=\linewidth]{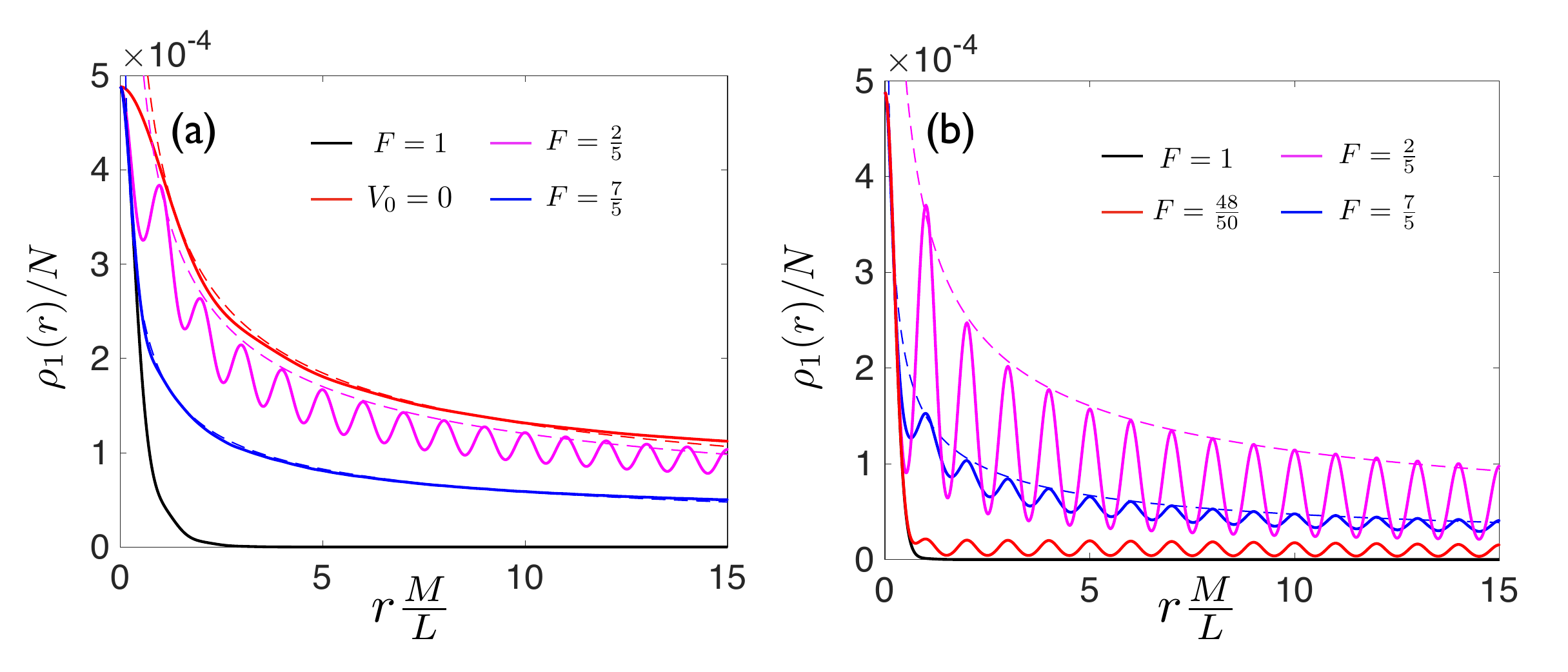} 
\caption{Auto-correlation functions as a function of $r$ for different filling factors and for a lattice depth of (a) $V_0 =5 E_r$ (except the red line) and (b) $V_0=20 E_r$. The dashed lines correspond to a $r^{-\frac{1}{2}}$ decay and are displayed for $V_0 =0$ and $F=\frac{2}{5},\frac{7}{5}$.}
 \label{fig:correlationfunctions} 
\end{figure}

This function then explicitly quantifies the long-range order of the system and we display it for different regimes in Fig.~\ref{fig:correlationfunctions}. One can see that for no external lattice (red curve in panel (a)), the correlations decay according to the well-known $r^{-\frac{1}{2}}$ power law behaviour of a free TG gas \cite{Cazalilla2011}. In the presence of a lattice for incommensurate particle density (such as $F=\frac{2}{5}$) an exponential decay is observed at small $r$ followed by a power-law decay. This demonstrates quasi-off-diagonal long range order (QODLRO) in the gas for the Hamiltonian \eref{eq:basicmodel} and mimics the behavior of the hard-core BHM \cite{Rigol2005}. In the continuous case, however, the power-law decay has an additional, very intuitive super-structure: the auto-correlations are modulated with the lattice periodicity. This corresponds to the fact that the atoms are preferably located at the lattice minima and their probability to be at a lattice maximum is significantly decreased. In fact, if the auto-correlations at these distances are going to zero, the system will enter a regime in which the tight-binding approximation works well. Once the second energy-band is occupied a deeper lattice is required to create modulations in the auto-correlations (compare $F=\frac{7}{5}$ in (a) and (b)).  For $F=1$ the pinned phase leads to an exponential decay of correlations, with no correlations surviving at distances longer than one lattice site for $V_0=20$. In the case of a microscopic number of defects ($F=\frac{48}{50}$) the initial exponential decay is much faster than for a macroscopic number of defects ($F=\frac{2}{5}$). In fact it is similar to the one found in the pinned phase, but unlike the pinned phase it shows small revivals at subsequent lattice sites. The magnitude of these revivals decays slightly, but it is difficult to ascertain the precise nature of this small decay, as we are restricted to a periodic 50-site lattice.

Since the modulations in the correlation-function stem from the off-diagonal terms in the RSPDM, they relate directly to the peaks that appear in the momentum distribution and therefore to the superfluid properties of the system. The external lattice breaks the continuious spatial symmetry and imposes discrete spatial symmetry, that is LRO (crystaline order), while the incommensurability between the number of lattice sites and particles allows for a superfluid flow giving rise to QODLRO. This results in numerous peaks at multiples of $ 2 k_l$ with a structure that is dependent on the lattice depth and particle-to-site ratio $F$. Contrarily, in the hard-core BHM for incommensurate lattice-sites and particle numbers, the momentum distribution shows only a single peak at $k=0$ associated with superfluidity. This is due to the spatial symmetry not being broken within the model. We note that a similar multi-peak structure for incommensurate systems is also obtained when a super-lattice is imposed on the BHM, as this also breaks the spatial symmetry and introduces oscillations in the correlation function \cite{Rigol2006}.

\section{Probing phases with a driven lattice}
\label{sec:driven}
Sudden perturbations are a powerful method for studying many-body effects in quantum systems, and for TG gases notable examples that have been investigated in recent years include the orthogonality catastrophe and decay \cite{Campo2011,Goold2011,Cetina2015, Parish2016, Demler2016, Demler2018}, persistent currents and probing superfluidity \cite{Citro2009, Hallwood2010,Schenke2011,Schenke2012,Schloss2016, Minguzzi2017,Minguzzi2017b,Minguzzi2018}. Such exploration of the non-equilibrium dynamics can be used to further characterise the different phases we have examined in the static model, and quenches can be used to detect phase transitions due to the creation of non-trivial dynamics induced by large quantum fluctuations at criticality \cite{Silva2008,Karrasch2013,Campbell2016,Heyl2017,Fogarty2017}.  

Here we quench by assuming that at $t=0$ our system is in an eigenstate at a given lattice depth and then suddenly rotating the external lattice potential at finite speed $v$, so that for $t>0$ we have $V_\mathrm{ext}(x_n-vt)=V_0 \cos^2(k_l (x_n-vt))$.  The time-dependent many-body wavefunction after the quench can still be calculated via the Bose-Fermi mapping theorem, so that we only need to evolve the initial non-rotating single particle states according to the time-dependent Schr{\"o}dinger equation 
\begin{equation}
i \hbar \partial_t \psi_n(x,t) = \left( -\frac{\hbar^2}{2 m} \frac{\partial^2}{\partial x^2}+V_\mathrm{ext}(x-vt) \right) \psi_n(x,t)\;.
\label{eq:timedependentequation}
\end{equation}
The time-dependent solutions can be found exactly in terms of the eigenfunctions $\psi_n(x,0)$ of the initial Hamiltonian and the eigenfunctions $\psi_{j,q}(x)$ and eigenvalues $E_{j,q}$ of the Hamiltonian in the co-rotating frame as \cite{Schenke2012} 
\begin{equation}
\psi_n(x,t) = e^{iqx}e^{-\frac{i \hbar q^2}{2m}t} \sum_j c_{jn} e^{-iE_{j,q}t/ \hbar} \psi_{j,q}(x-vt),
\label{eq:timedependentsol}
\end{equation} 
where $q=\frac{m v}{\hbar}$. The eigenfunctions in the co-rotating frame are a solution to the Schr\"odinger equation 
\begin{equation}
E_{j,q}\psi_{j,q}(x) = \left( -\frac{\hbar^2}{2 m}\frac{\partial^2}{\partial x^2}+V_\mathrm{ext}(x) \right) \psi_{j,q}(x),
\label{eq:comovingeq}
\end{equation}
obeying the twisted boundary conditions $\psi_{j,q}(x+L)=e^{-i q L}\psi_{j,q}(x)$. The coefficients $c_{jn}$ then contain all the information about the initial state of the system and are defined as
\begin{equation}
   c_{jn}= \langle \psi_{j,q}(x)|e^{-iqx}| \psi_n(x,0) \rangle = \int dx \, \psi_{j,q}^*(x)e^{-iqx} \psi_n (x,0).
\end{equation}

The applied rotation will lead to transport of particles with respect to the lab frame and the average flow of particles can be quantified by calculating the average momentum $K(t) = \int n(k,t)\, k\, dk$. While calculating the full momentum distribution is a numerically demanding process for large particle numbers, the average momentum can be calculated efficiently as it is related to the probability current density of the system as $  K(t) =  \frac{1}{L}  \int j(x,t)\;dx\;$. For a TG gas, the probability current density is identical to that of the free Fermi-gas and is simply given as the sum of the single-particle currents corresponding to the occupied energy levels
\begin{equation}
  j(x,t) = \frac{\hbar}{m} \mathrm{Im} \sum_{n=0}^{N-1} \psi^*_n(x,t)\, \partial_x \psi_n(x,t).
\end{equation} 
For the time-dependent wavefunction given in Eq.~\eref{eq:timedependentsol}, the explicit form of the average momentum can therefore be calculated as
\begin{eqnarray}
K(t) &=&K_0 +K_t(t) \nonumber \\
 &=&\frac{N \hbar q}{m L} + \frac{\hbar}{m L} \sum_{n=0}^{N-1}  \sum_{j} |c_{jn}|^2 \mathrm{Im} \left[ F_{jj}\right] \nonumber \\ 
&+&\mathrm{Im} \left[\frac{\hbar}{m L} \sum_{n=0}^{N-1} \sum_{j \neq k} c_{jn} c_{kn}^* e^{-i(E_{k,q}-E_{j,q})t/ \hbar}  F_{jk}(t)\right],
\label{eq:explicitcurrent}
\end{eqnarray}
where 
\begin{equation}
F_{jk}(t) = \int dx  \psi_{j,q}^*(x-vt) \, \partial_x \psi_{k,q}(x-vt).
\label{Fjk}
\end{equation}
The elements $F_{jj}$ are independent of time in all cases we consider, as the two eigenfunctions are the same, which means that the time-dependence just shifts the integrand. For $j \neq k$, however, the two eigenfunctions are different and the time-dependence of the integrand is therefore more complicated and does affect the value of the integral. The average momentum therefore has a time-independent part $K_0$ consisting of the first two terms in  Eq.\eref{eq:explicitcurrent} and a time-dependent part $K_t(t)$ which consists of the last term.

The effect of the underlying many-body phase of the TG gas on the average transport in the system should be reflected in the average momentum and Eq.~\eref{eq:explicitcurrent} allows us to explain the observed behavior in some mathematical detail. Furthermore, the same behaviour appears for a spin-polarised Fermi gas, even though its momentum distribution is entirely different. This is because the expectation value of the momentum operator is the same for the Fermi and TG gas, which means that the average momentum alone is insufficient for fully characterising the dynamics of the different many-body bosonic phases. We will therefore investigate and compare the average momentum, the coherence and the momentum distribution.

\subsection{Preliminary consideration - rotation speed}
\label{subsubsec:rotationspeed}
\begin{figure}[tb] 
\centering
\includegraphics[width=\linewidth]{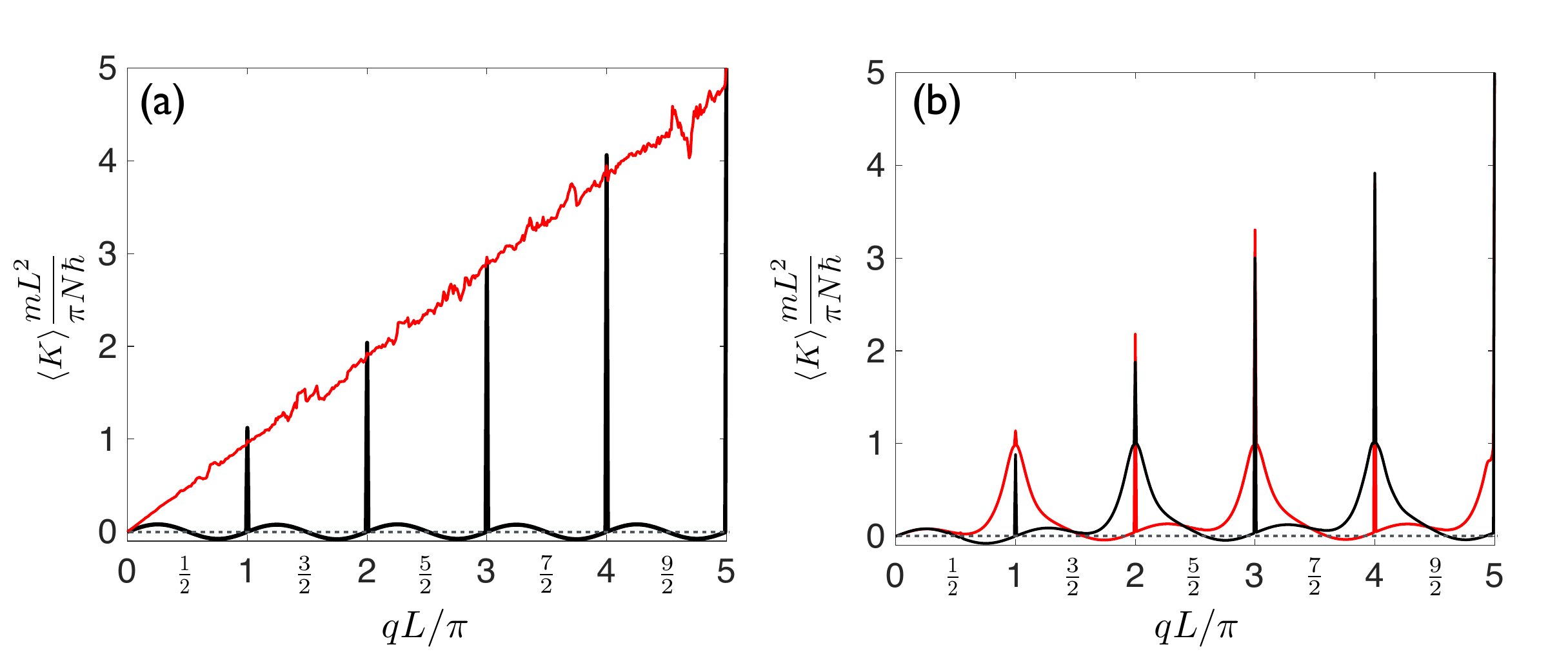}
\caption{$\langle K \rangle/N$ as a function of $q$ for different filling ratios and depths. In (a) the black line corresponds to $F=\frac{2}{5}$ and $V_0=0.02 E_r$, while the red line corresponds to  $F=\frac{2}{5}$ and $V_0= 20 E_r$. In (b) the black line corresponds to $F=\frac{49}{50}$ and $V_0=0.02 E_r$, while the red line corresponds to $F=1$ and $V_0=0.02 E_r$. The dashed black line indicates $\langle K \rangle/N=0$.} 
 \label{fig:currentasfunctionofspeed} 
\end{figure}

\begin{figure}[tb] 
\centering
\includegraphics[width=\linewidth]{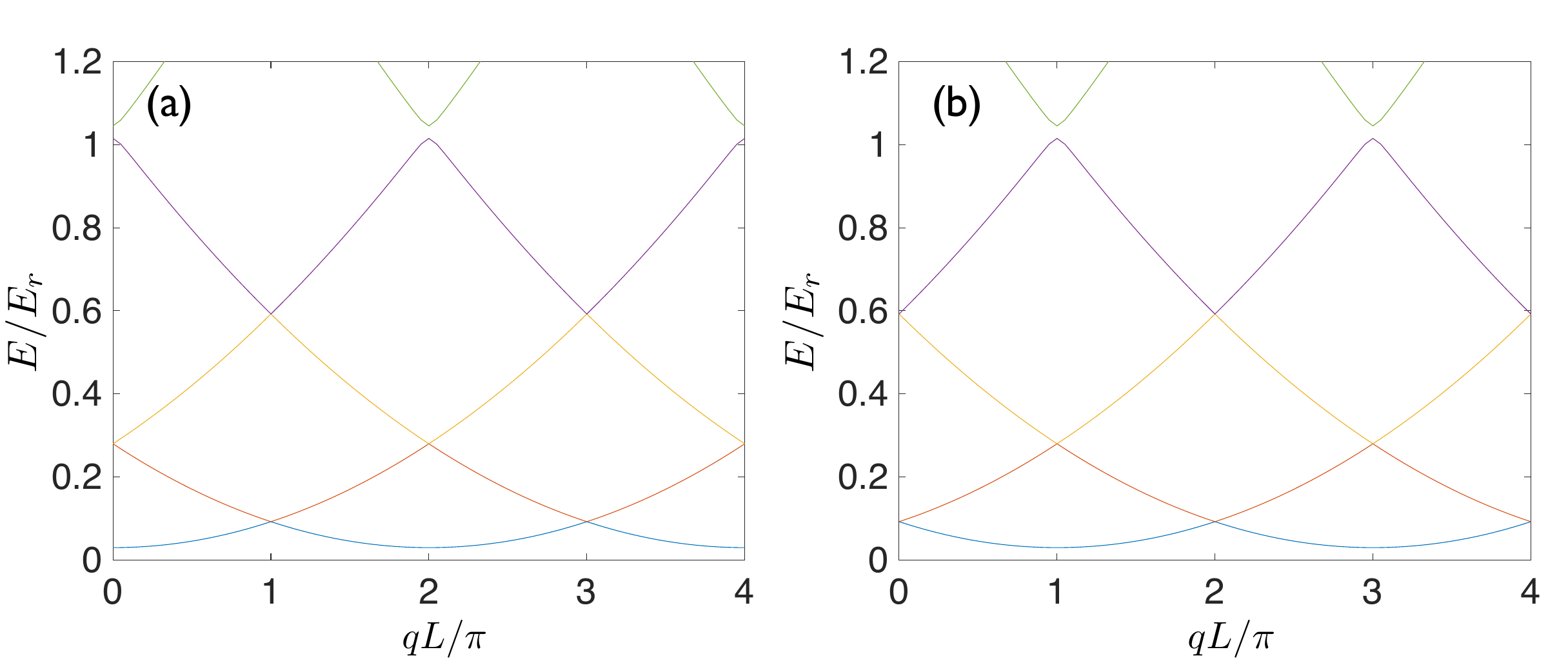}
\caption{(a) Single-particle energy spectrum for PBC (odd number of particles)  and (b) A-PBC (even number of particles), plotted as a function of the driving momentum $q$ for $M=4$ and $V_0=0.06 E_r$.} 
 \label{fig:oddevenspectrum} 
\end{figure}

It was shown by Schenke {\it et al.} \cite{Schenke2012} that persistent currents can be created in a TG gas by driving it with a rotating delta-barrier. If the driving momentum is $q = \eta \pi / L$, where $\eta$ is an integer, this yields maxima in the time-averaged average momentum per particle $\langle K \rangle/N$, while outside these values it stays essentially zero. A similar resonant behaviour is also present in our system when rotating the lattice potential at small lattice depths and for incommensurate filling (see black curve in Fig.~\ref{fig:currentasfunctionofspeed}), however, the momentum no longer vanishes away from the resonance values. In fact it oscillates between positive and negative momentum with a small amplitude. In order to simplify the discussion in this section we define the scaled driving momentum $\Omega=q L /\pi$. For $F=1$ the values of $\langle K \rangle/N$ around $\Omega= 2 \eta+1$ become large, while the same behavior can be seen for $F=\frac{49}{50}$ around $\Omega= 2 \eta$. This can be understood by looking at the single-particle spectrum for odd particle numbers (with PBC) and for even particle numbers (with A-PBC) as a function of $q$, which is plotted in Fig.~\ref{fig:oddevenspectrum} for $M=4$ and $V_0 = 0.06 E_r$. Note that these numbers are chosen for visual simplicity, and the structure is exactly the same for $M=50$ and $V_0 = 0.02$ (with a slightly smaller gap). The structure of the two spectra reflects the difference in the average momentum, that is the spectrum for odd particle numbers close to $\Omega=2 \eta+1$ looks similar to the one for even particle numbers close to $\Omega=2 \eta$. The reason for this difference is that the ground-state is non-degenerate in the initial state for an odd number of particles, while it is two-fold degenerate in the initial state for an even number of particles. A lattice with an even number of sites therefore forces a splitting of the $M,M+1$ degenerate states for odd particle numbers, but there is no splitting for even particle numbers. An odd number of sites switches around the behavior for odd and even particle numbers, but the explanation is analogous. For deeper lattices we see that $\langle K \rangle/N$  grows linearly with $q$ even for incommensurate particle densities. A detailed explanation of the behavior in the different regimes shown in Fig.~\ref{fig:currentasfunctionofspeed} can be found in Appendix A.

\subsection{Many-body phases in the driven system}
As we are interested in using the driven lattice to probe the phases of the system, we restrict our simulations to small values of the rotation speed, namely $q = 2.9 \pi/ L$, which is off resonant so that the rotation does not excite the particles into higher bands. This therefore allows for a clear distinction between the dynamics of the superfluid and insulating phases in the system. In order to understand the nature of the bosonic phases, we investigate the coherence and the momentum distribution in addition to the average momentum.

\subsubsection{Time-averaged Dynamics:}
\label{subsubsec:timeaveraged}
\begin{figure}[tb]
\centering
\includegraphics[width=\linewidth]{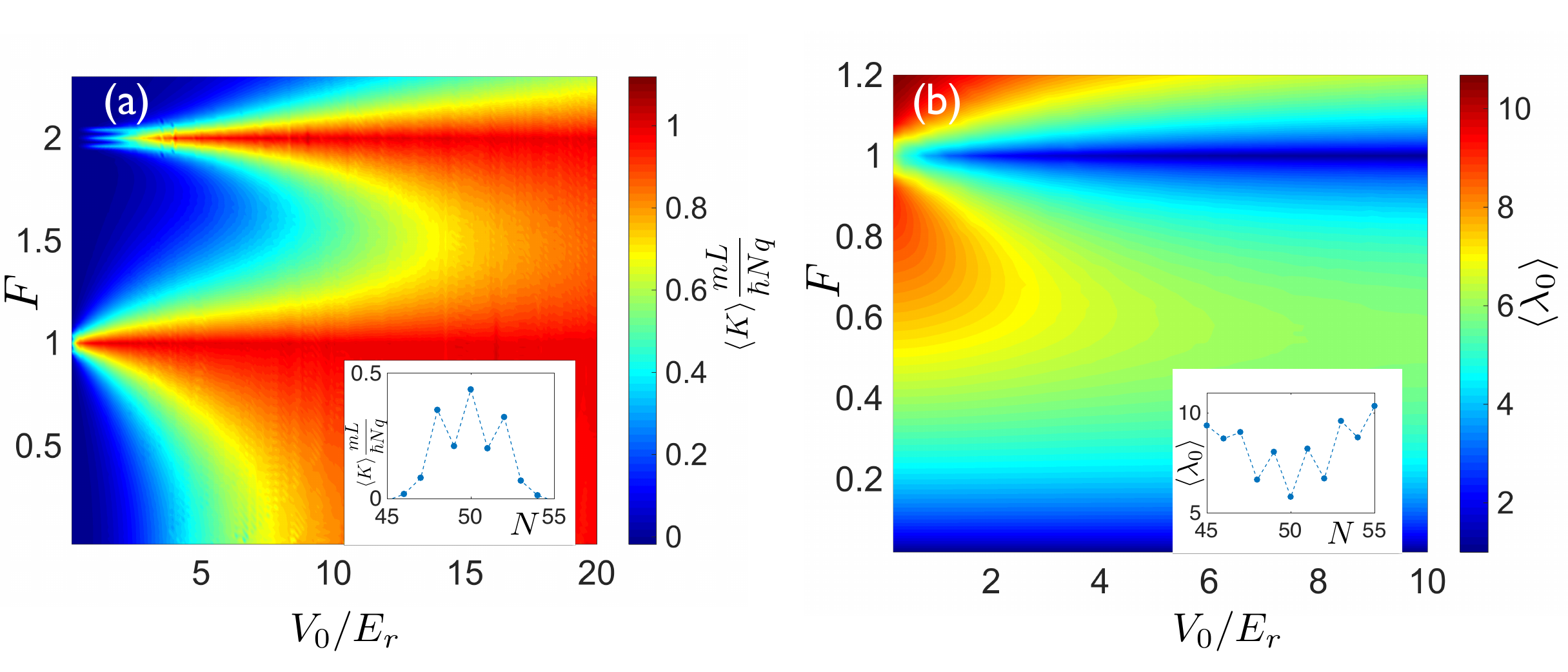}
\caption{(a) Time-averaged average momentum and (b) time-averaged coherence as a function of the lattice depth $V_0$ and the filling ratio $F$. The insets shows the same near $F=1$ for a lattice depth of $V_0=0.1 E_r$. These plots are based on calculations for $M=50$ and with the time-average taken over the interval 0 to 10$t_0$. }
 \label{fig:dynamicphasediagram} 
\end{figure}

In Fig.~\ref{fig:dynamicphasediagram} we show the time-averaged average momentum  $\langle K \rangle/N$ and the time averaged coherence $\langle \lambda_0 \rangle$ as a function of the filling ratio, $F$, and the lattice depth, $V_0$. For $\langle \lambda_0\rangle$ one can see qualitatively the same behavior as for the static coherence shown in Fig.~\ref{fig:staticphasediagram}(a), although for $\langle \lambda_0 \rangle$ after the quench an odd-even effect is present near commensurability in shallow lattices (see the inset in Fig.~\ref{fig:dynamicphasediagram}(b)). The same odd-even effect is present for $\langle K \rangle/N$, as discussed in section \ref{subsubsec:rotationspeed}, and for which a detailed explanation of the origin can be found in Appendix A. The minimum in the coherence obtained for $F=1$ indicates the presence of the pinned phase and the maximum in $\langle K \rangle/N$ at $F=1$ (and $F=2$) can therefore be attributed to this state, which restricts tunneling between the sites resulting in the gas being moved along with the lattice. For incommensurate values of $F$, the relatively large degree of coherence indicates the presence of a superfluid phase (as does the small value of  $\langle K \rangle/N$), where particles can freely tunnel through the lattice as it rotates through the gas. 

\begin{figure}[tb]
\centering
\includegraphics[width=\linewidth]{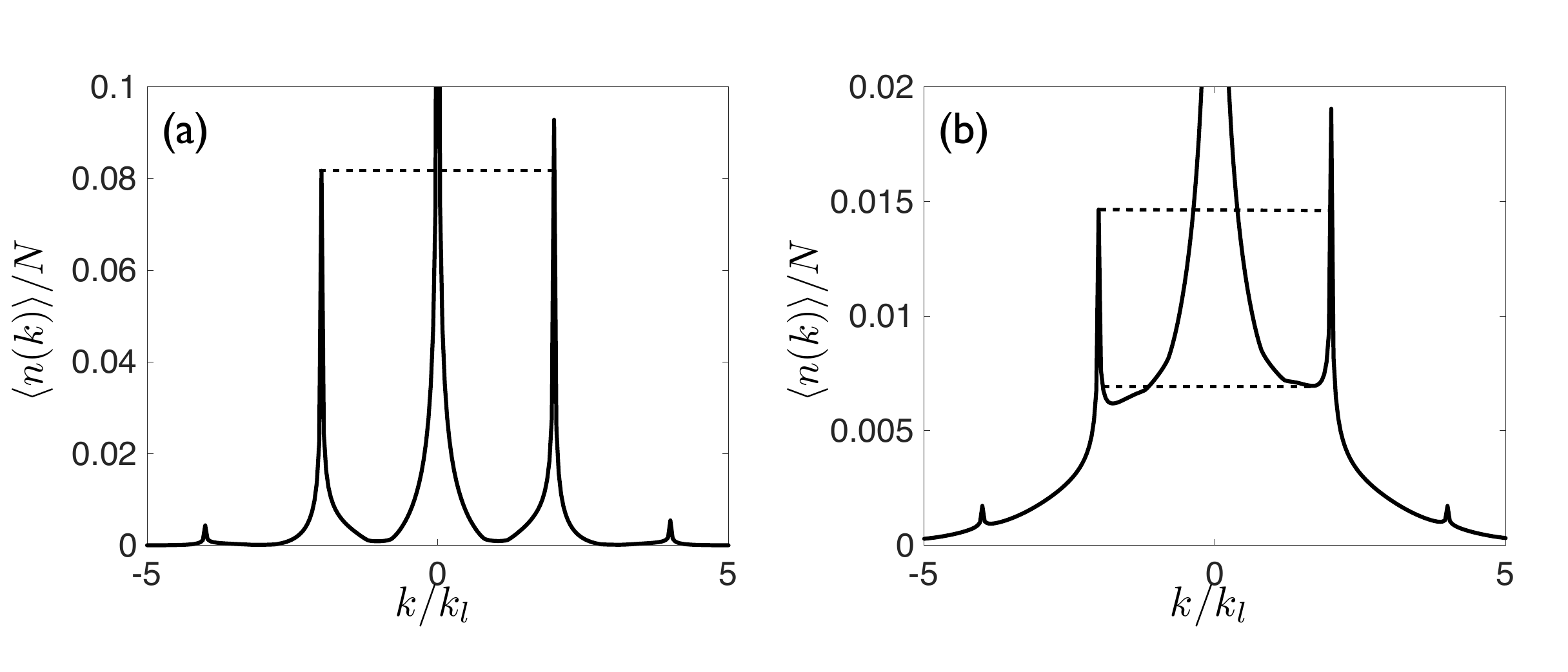}
\caption{Time-averaged momentum distribution (black line) after a rotational quench for $F=\frac{2}{5}$ (a) and $F=\frac{7}{5}$ (b). The dashed black lines indicate the asymmetries between positive and negative momentum occupations.} 
 \label{fig:timeaveragedmomentumdistribution} 
\end{figure}

For deep lattices ($V_0>10E_r$) we find $\langle K \rangle / N \approx qL/ \pi$ for both commensurate and incommensurate fillings and this quantity thus no longer distinguishes between the commensurate insulating and the incommensurate supersolid-like phases. The time-averaged coherence, however, clearly distinguishes between these phases which implies that the time-averaged full momentum distribution must contain information about their dynamical properties as well. For $F=1$ the pinned phase Gaussian-like momentum distribution is slightly asymmetrical, leading to the overall transport observed, while the momentum distribution in the frictionless superfluid phase (for $V_0\ll E_r$ and $F \neq \mathcal{N}$)  has a single, symmetrical superfluid peak centered at $k=0$, even when rotation is applied. As this is difficult to visually distinguish we do not plot the time-averaged momentum distributions in these cases.  However, the time-averaged momentum distributions after a quench for $F=\frac{2}{5}$ and $F=\frac{7}{5}$ are shown in Fig.~\ref{fig:timeaveragedmomentumdistribution}.  For $F=\frac{2}{5}$, corresponding to a macroscopic number of defects, the transport is obtained due to an asymmetry in the population of the back-scattering momentum peaks, with the peaks at positive momenta having a higher probability than the peaks at negative momenta. For $F=\frac{7}{5}$  a combination of an asymmetrical Gaussian shape (for the first 50 pinned particles) and the asymmetrical back-scattering momentum peaks (for the remaining 20) is responsible for the transport. In the case of a microscopic number of defects the bulk of the transport is due to the pinned phase, but a small part is contributed from the asymmetry between the population of the back-scattering momentum peaks, similar to what is seen for a macroscopic number of defects in Fig.~\ref{fig:timeaveragedmomentumdistribution}. So although $\langle K \rangle / N$ is the same for the pinned phase and the supersolid-like phase at different filling ratios in the deep lattice, the time-averaged momentum distributions and the type of transport is very different, reflecting the different natures of the two many-body phases.

\subsubsection{Instantaneous dynamics:}
\begin{figure}[tb]
\centering
\includegraphics[width=\linewidth]{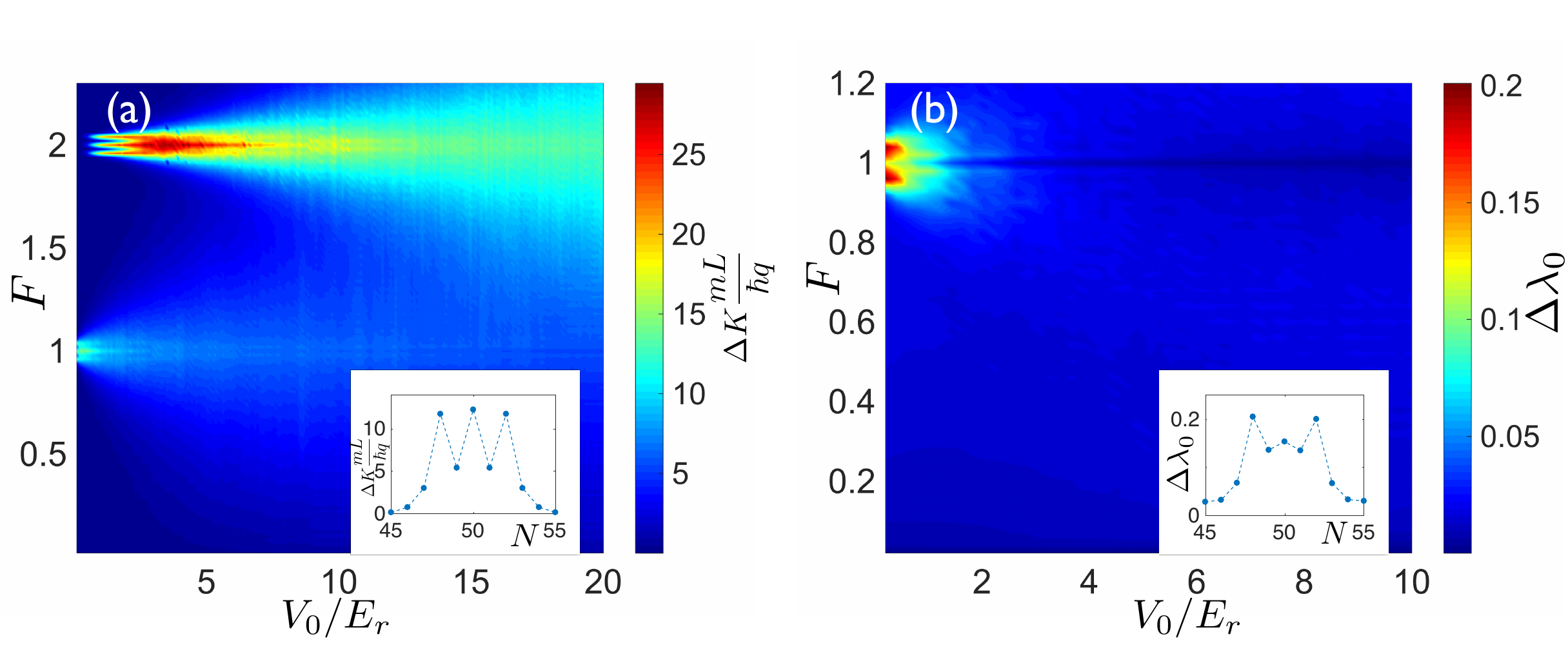}
\caption{ Standard deviation of (a) the average momentum and (b) the time-averaged coherence as a function of the lattice depth and the filling ratio. The insets shows the same near $F=1$ for a lattice depth of $V_0=0.1 E_r$. These plots are based on calculations for $M=50$ and with the time-average taken over the interval 0 to 10$t_0$.} 
 \label{fig:fluctuationsofdynamicphasediagram} 
\end{figure}

The time-averaged variables only give a partial picture of the dynamics and a fuller understanding can be gained by considering the instantaneous properties of the average momentum and coherence. For this we plot the standard deviation of the average momentum and coherence in Fig.~\ref{fig:fluctuationsofdynamicphasediagram} (based on the same calculation as the time averages reported in Fig.~\ref{fig:dynamicphasediagram}). These can be seen to be essentially zero in the superfluid phase for shallow lattices (incommensurate particle numbers), while they become large as $F$ approaches integer values. For deeper lattices the standard deviation of the coherence stays zero, while the standard deviation of the average momentum becomes larger. To understand the physics under-lying these fluctuations in more detail, we will next consider the instantaneous properties of the coherence and average momentum at some representative values of the depth and filling ratio.

\begin{figure}[tb]
\centering
\includegraphics[width=\linewidth]{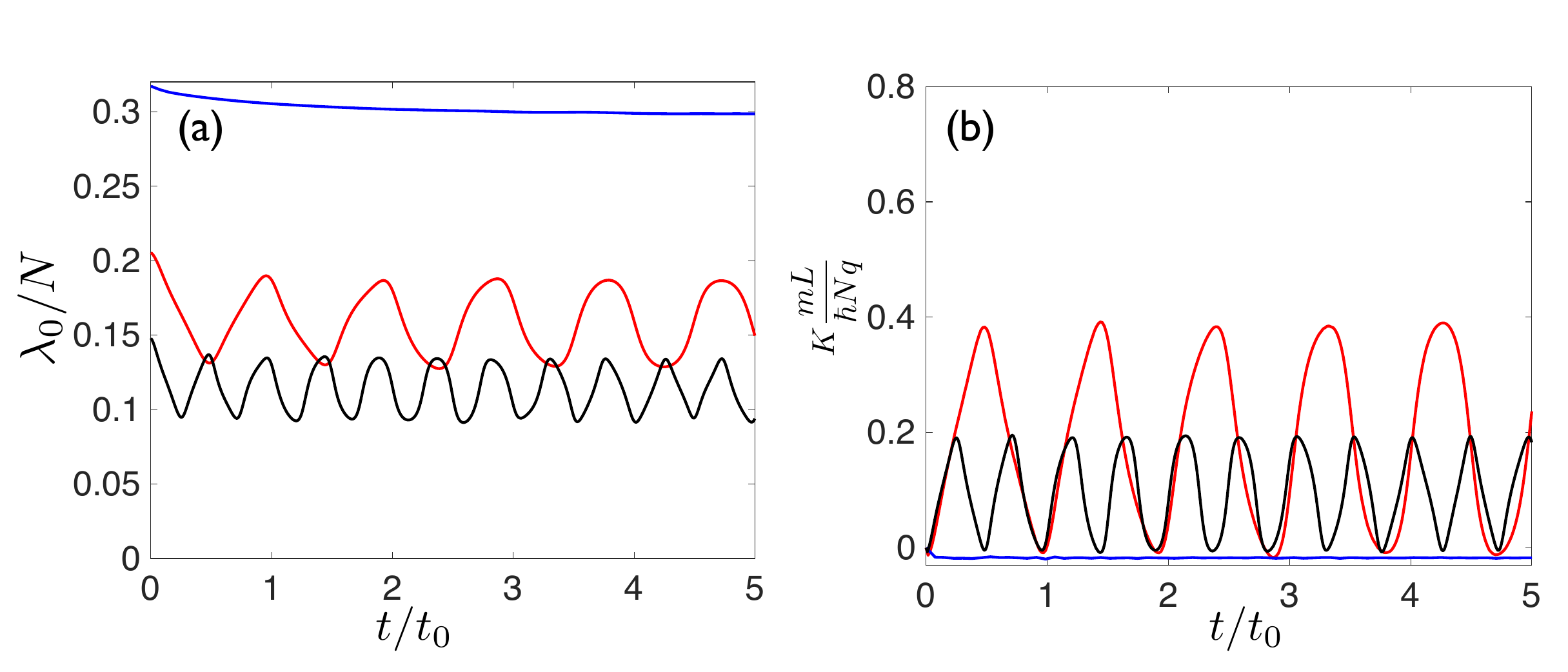}
\caption{(a) Coherence $\lambda_0/N$ and (b) average momentum per particle as a function of time after a quench. The red lines correspond to $V_0 =0.02E_r$ and $F=1$, while the blue lines correspond to $V_0 =0.02 E_r$  and $F=\frac{2}{5}$. The black lines correspond to $V_0 = 0.8 E_r$ and $F=2$.} 
 \label{fig:instantaneousplotsshallowlattice} 
\end{figure}

The dynamics of the coherence, $\lambda_0(t)/N$, and of the average momentum, $K(t)$, after the quench are shown in the shallow lattice limit in Fig.~\ref{fig:instantaneousplotsshallowlattice}.  In the commensurate case (red line) the gas exhibits collective many-body oscillations between the coherent superfluid and a somewhat less coherent insulating phase with a periodicity $\sim t_0 =L/(Mv)$. The same behaviour appears for $F = 2$ (black line), but the frequency of the oscillation has roughly doubled. The average momentum in the insulating phase is much higher than the superfluid phase, which means that the transport is very similar to classical stick-slip motion described by the Frenkel-Kontorova model  \cite{Braun2013}. The periodicity of these oscillations is derived in Appendix A and shown to be related to the discrete time-symmetry of the Hamiltonian for the average momentum. This type of collective oscillation is present for both the coherence (a many-body measure) and the average momentum (single-particle measure), whereby the momentum dynamics is clearly a reflection of the oscillation between many-body phases as implied by the coherence. For $F =\frac{2}{5}$, where a macroscopic number of defects exist, time-fluctuations in any of the parameters are essentially absent and the gas is always in the coherent superfluid phase which corresponds to the average momentum being small and slightly negative, due to our choice of $q$ (see section \ref{subsubsec:rotationspeed} and Appendix A) at all times. The existence of the superfluid phase thus results in essentially frictionless dynamics, in which particles simply tunnel through the shallow barriers without responding to them. In shallow lattices, for both commensurate fillings ($F=1$) and microscopic numbers of defects, there exist regular oscillations of the coherence and average momentum. As the depth is increased, but still within the region of non-zero fluctuations of the coherence (see Fig.~\ref{fig:fluctuationsofdynamicphasediagram}(b)), the stick-slip motion becomes less regular. The collective many-body fluctuations around $F=1$ in shallow lattices is suggestive as this region corresponds to the critical region for the commensurate-incommensurate pinning transition. The large fluctuations observed in the coherence, which is an order parameter that distuingushes the pinned and superfluid phases, is therefore a manifestation of the underlying incommensurate-commensurate transition. This aligns our results with other indications that critical points and regions introduce large fluctuations in the dynamics of order-parameters \cite{Sachdev,Kinross2014,Mukh2012,Witczak2014,Lelas2012,Campbell2016,Heyl2017}.

\begin{figure}[tb]
\centering
\includegraphics[width=\linewidth]{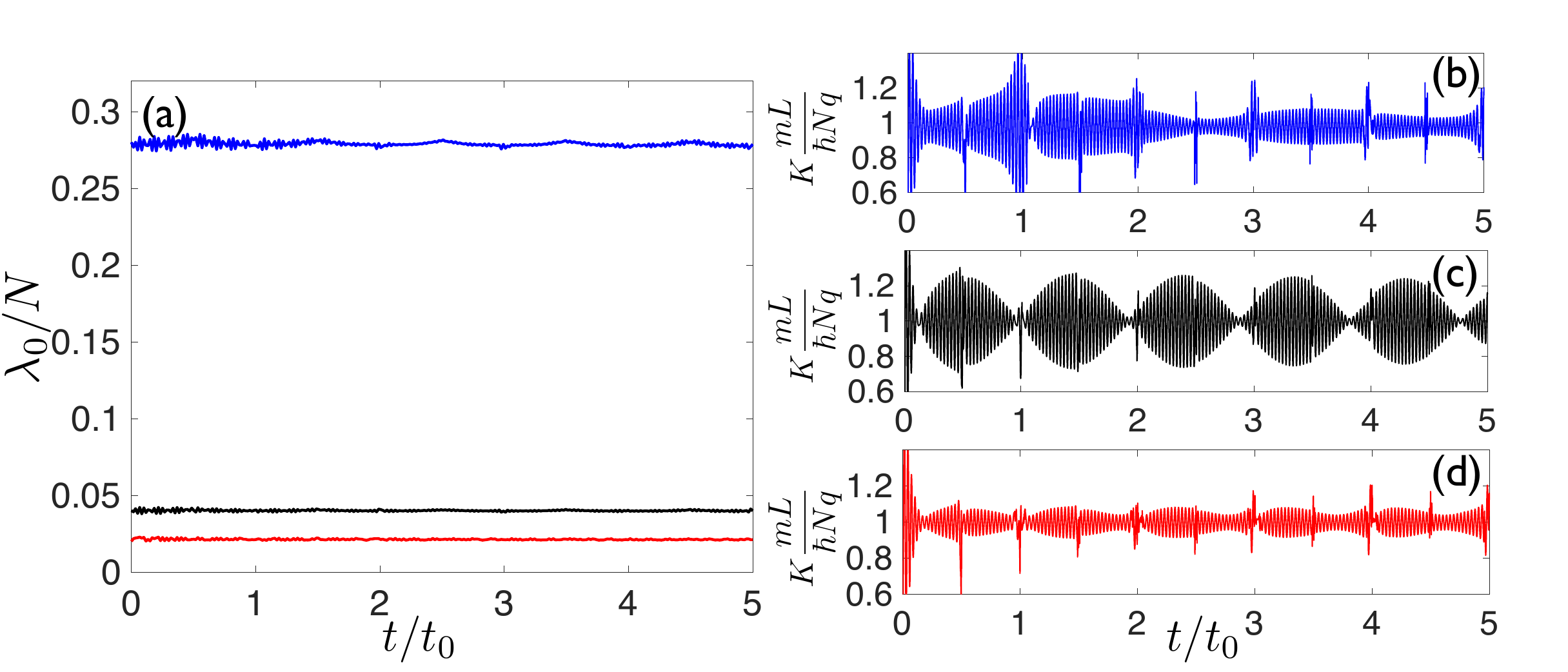}
\caption{(a) Coherence $\lambda_0/N$ and (b)-(d) average momentum per particle as a function of time for $V_0 = 20 E_r$. The red lines correspond to $F=1$, while the blue lines correspond to $F=\frac{2}{5}$ and the black lines correspond to $F=\frac{49}{50}$.} 
 \label{fig:instantaneousplotsdeeplattice} 
\end{figure}

For deeper lattices the fluctuations in the coherence are small (see Fig.~\ref{fig:fluctuationsofdynamicphasediagram}(b)) and the instantaneous coherence is therefore not expected to contain any information that cannot be understood from the average coherence, although we plot it for completeness in Fig.~\ref{fig:instantaneousplotsdeeplattice}(a). Indeed, one can see that there are only small fluctuations and that the coherence is close to 1 in the commensurate pinned phase, while the gas is slightly more coherent in the presence of a microscopic number of defects ($F=\frac{49}{50}$) and approaches a $\sqrt{N}$ degree of coherence for a macroscopic number of defects ($F=\frac{2}{5}$). This directly reflects the results shown in Fig.~\ref{fig:dynamicphasediagram}(b). For the average momentum, however, Fig.~\ref{fig:instantaneousplotsdeeplattice}(b) shows that large fluctuations are still present in deep lattices. As these fluctuations are only present for the average momentum they do not correspond to oscillations between many-body phases as was the case for the shallow lattice oscillations. In this deep lattice limit particles are transported along with the lattice on average with $\langle K \rangle = qL/ \pi$ and the short-time oscillations with period $ \sim \frac{2 \pi \hbar}{E_{M+1}-E_{M}}$ correspond to particles being excited to higher bands at each lattice site, creating on-site dynamics due to the finite width of the continuum lattice (see Appendix A). These types of fluctuations are clearly distinct from the fluctuations between many-body phases observed in shallow lattices. Additionally they are present for both commensurate and incommensurate particle densities. The instantaneous average momentum is therefore not a useful way to distinguish between the pinned phase and the supersolid-like phase in the deep lattice limit. For that information an investigation of the full momentum distribution is required. As can be understood from the coherence, however, the time fluctuations of the full momentum distribution are not particularly interesting: the time-averaged momentum distribution, which we discussed in section \ref{subsubsec:timeaveraged}, already contains all the relevant information.  

\section{Conclusion}
\label{sec:conclusions}
Utilizing the Bose-Fermi mapping theorem and studying the reduced single-particle density matrix we have investigated the phases of the Tonks-Girardeau gas in a continuum optical lattice model for commensurate and incommensurate particle densities. For shallow lattices we reproduce the well-known results, namely the existence of a pinned insulating phase for commensurate particle densities and a superfluid-like phase for incommensurate particle densities which possesses a similar degree of coherence as the free TG gas. The momentum distribution reflects these phases; in the pinned insulating phase it is Gaussian-like and delocalised, while it is centered around a single peak at $k=0$ for incommensurate particle numbers. For deeper lattices, we find that a supersolid-like phase emerges for incommensurate particle numbers. This defect-induced superfluid phase is well-characterized by the momentum distribution, which shows a multi-peaked structure reflecting spatial modulations in the auto-correlation in the presence of an optical lattice due to the breaking of the continuous spatial symmetry. 

The momentum distribution depends sensitively on the filling ratio and it is useful to consider two distinct, but continuously connected regimes; a microscopic number of defects where we have shown that the momentum distribution corresponds to a large number of pinned particles, with a few superfluid particles (or holes) on top and a macroscopic number of defects, where the multi-peak structure is dominant. Additionally we have shown that dynamically rotating the lattice allows one to probe the structure of the phases. The average momentum, which measures the degree of transport in the system, clearly reflects the shallow lattice phase diagram with maximum transport obtained for commensurate particle densities due to the pinned phase. Additionally a collective many-body stick-slip-like behavior is observed for particle numbers close to commensurability where the gas oscillates between the pinned and superfluid phases with a period $T \approx t_0$. On the other hand, in deeper lattices, despite the supersolid-like phase retaining some coherence, the average particle transport is very similar to the pinned phase, as the particles are transported along with the lattice on average. The average momentum has short time-scale oscillations in this regime corresponding to on-site dynamics induced by on-site particle excitations in the lattice. These oscillations are therefore very different from the many-body phase oscillation observed for shallow lattices. Investigating the full momentum distribution, however, shows that the underlying physics of the transport is very different for commensurate and incommensurate particle densities, the former is the result of an asymmetric Gaussian-like shape, while the latter is the result of an asymmetry between the occupation of the positive and negative back-scattering momentum side-peaks.

Our approach is experimentally realizable as ring lattice potentials have been created using spatial light modulators \cite{Dumke,Dumke2016}, and alternative possible setups include optical nanofibers \cite{Phelan2013} and rapidly moving lasers that can ``paint'' arbitrary optical trap shapes \cite{Henderson2009}. Another possibility is using an optical lattice in a long one-dimensional box trap. Theoretically, the calculation of such dynamics for a TG-gas on a lattice confined by a square well is feasible and an obvious future extension of the work presented in this paper. Another natural extension is to consider finite interactions, rather than the TG limit. This would allow for an investigation of the interplay between interactions and commensurability, similar to the results presented in \cite{Brouzos2010}, but in the presence of a moving lattice. Solving the dynamics of such systems is feasible for small particle numbers, for example by utilizing the exact diagonalization scheme outlined in \cite{Deuretzbacher2007}. 

\ack
The authors thank Rashi Sachdeva for frictionless discussions. This work was supported by the Okinawa Institute of Science and Technology Graduate University. TF acknowledges support under JSPS KAKENHI-18K13507. 

\appendix 
\section{Detailed explanation of the average momentum behavior after a quench}
\label{sec:app}
\setcounter{section}{1}
In this appendix we present a detailed analysis of Eq.~\eref{eq:explicitcurrent} in the different regimes of interest, which will allow for a mathematical explanation and clarification of the observed behavior of the average momentum. This in turn helps clarify the physical properties and processes responsible for this behaviour. The appendix is structured as follows: The first section explains the incommensurate behavior in the shallow lattice as a function of the driving momentum $q$, by considering the free space wavefunctions analytically. The second section contains a numeric investigation of  the important quantities in Eq.~\eref{eq:explicitcurrent} in the presence of a lattice potential. 

\subsection{Free space analysis}
The incommensurate gas in shallow lattices can be qualitatively understood by evaluating the expression for the average momentum given in  Eq.~\eref{eq:explicitcurrent} for plane-waves, assuming that the lattice only leads to a small perturbation. In free space and with PBC (the analysis for A-PBC for even particle numbers is similar) we can therefore assume the wavefunctions to be given by 
\begin{equation}
   \psi_l(x,0) = \frac{1}{\sqrt{L}} e^{ i 2 l \pi x/L} \quad  , \quad \psi_{j,\Omega}(x) = \frac{1}{\sqrt{L}} e^{i(\Omega+2 j) \pi x/L},
\end{equation} 
where $l,j=\lbrace 0,\pm 1,\pm 2,\pm 3 \dots\rbrace$. The energies in the rotating frame are therefore given by $E_j = \frac{\hbar^2 \pi^2}{2m L^2}(\Omega+2j)^2$.  The  points $\Omega=\kappa$ (where $\kappa$ is an integer) are special as the rotational eigenstates $\psi_{j,\kappa}(x)$ and $\psi_{j',\kappa}(x)$  are degenerate for $j'=-j-\kappa$. For $\kappa=2 \eta $ we have $j'=j$ with $j=-\eta$, therefore the lowest energy state will be unpaired, see Fig.~\ref{fig:oddevenspectrum}(a). This means that almost all eigensfunctions of the system can be expressed alternatively as entirely real or imaginary linear combinations of the $j$ and $j'$ eigenstates  of the form $\phi_{j+,\kappa}(x)=\sqrt{\frac{2}{L}} \cos(\pi(\kappa+2j)x/L)$ and $\phi_{j-,\kappa}(x)=\sqrt{\frac{2}{L}} i\sin(\pi(\kappa+2j)x/L)$. This then leads to $\mathrm{Im}(F_{jj}(t))=0$, which means that $K(t)=\frac{N \hbar q}{m L}+K_t(t)$. The time-dependent-part of the average momentum creates oscillations around the time-independent part, which matches exactly the average visible in Fig.~\ref{fig:currentasfunctionofspeed}. 

For values $\Omega \neq \kappa $ we find
\begin{eqnarray}
  c_{jl} &=&  \frac{\sin(\pi[-\Omega - j + l]))}{\pi (-\Omega - j + l)},  \\
  F_{jk}(t) &=& i[\Omega+2 k]\pi/L e^{i 2 \pi(j-k) vt/L} \delta_{jk},
\end{eqnarray}
which means that there is no time-dependence of the average momentum within this approximation. 
This is shown to be approximately true in Fig.~\ref{fig:instantaneousplotsshallowlattice} for the shallow lattice limit as well. For values $\Omega = \kappa/2 $, we find $|c_{jl}|^2=|c_{j'l}|^2$ when $j'=2l-n-j$,  which can be used to evaluate the average momentum analytically as
\begin{eqnarray}
K(t) &=& \frac{N \hbar q}{m L} + \frac{\hbar}{ m L} \sum_{l=-(N-1)/2}^{(N-1)/2}  \sum_{j} |c_{jl}|^2 [q+2 \pi j/L]  \\
&=& \frac{2 N \hbar q}{m L} + \frac{\hbar}{2 m L^2} \sum_{l=-(N-1)/2}^{(N-1)/2}   \sum_{j} |c_{jl}|^2 2 \pi[2l-n] \\
&=& \frac{2 N \hbar q}{m L} + \frac{-N 2 \pi n \hbar}{2 m L^2} =  \frac{2 N \hbar q}{m L} - \frac{2 N \hbar q}{ m L} = 0.
\end{eqnarray}
This corresponds to the average value we obtain for the full numerical calculations shown in Fig.~\ref{fig:currentasfunctionofspeed}. For $\kappa < \Omega < \kappa / 2 $ we have $\left|  \sum_{l=-(N-1)/2}^{(N-1)/2}\sum_{j} |c_{jn}|^2 \mathrm{Im} [F_{jj}]\right| < N q$,  while for $\kappa / 2  < \Omega < \kappa$ we find $\left|\sum_{l=-(N-1)/2}^{(N-1)/2}\sum_{j} |c_{jn}|^2 \mathrm{Im} [F_{jj}]\right| >N q$. This gives rise to the regions of negative average momentum visible in in Fig.~\ref{fig:currentasfunctionofspeed}, in which particles rotate in the direction opposite to the lattice. As $\Omega = \kappa $ is approached the average momentum tends towards zero, before the discontinuous jump to $\langle K(t) \rangle=\frac{N \hbar q}{m L}$ at the point where the energy spectrum becomes degenerate. 

\subsection{Numeric analysis for the lattice}
To explain the behavior close to commensurability and for deeper lattices, we use the numeric wavefunctions obtained by finite difference diagonalization. The introduction of the energy gap changes the relative amplitude between the real and imaginary parts of $\psi_{M-d,\Omega}(x)$ ($d=0,1...$), with an opposite shift in $\psi_{M+1+d,\Omega}(x)$ close to $\Omega=2 \eta$ for odd particle numbers. Similar shifts are observed close to $\Omega=2 \eta+1$, but they are negligible for very small gaps ($V_0 = 0.02 E_r$), although the difference becomes less pronounced as the gap size is increased. As expected, the opposite behavior with respect to $\Omega$ is observed for the systems with even particle numbers. 

In the free space analysis $F_{jk}(t) = 0$ for $j \neq k$, which is still approximately true once the lattice is introduced, but as a consequence of the change in the wave-functions the terms $F_{(M-d)(M+1+d)}(t) = F_{(M+1+d)(M-d)}(t) \neq 0$ contribute significantly to the time-oscillating part of the average momentum. $F_{(M-d)(M+1+d)}(t)$ generally oscillates with a periodicity related to  $t_0=L/(Mv)$ as the discrete time symmetry of the system means that the single-particle wave functions obey the relation $\psi_{j,\Omega}(x-vt)=\psi_{j,\Omega}(x-v(t+t_0))$. The size of the $d>0$ contributions are much smaller than the $d=0$ contribution for shallow lattices, but as the lattice depth increases so does the size of $ F_{(M-d)(M+1+d)}$. 

\subsubsection{Commensurable particle numbers in the shallow lattice}
For very small gaps $(V_0 = 0.02 E_r)$ only the $d=0$ contributions matter and we will consider this case in some detail in order to explain the behavior close to commensurability. This contribution is only important when $c_{Mn}^{\vphantom{-}} c_{(M+1)n}^{*} \neq 0$, that is when the overlap between the initial state and the $M$ and $M+1$ rotating states are comparatively large. For A-PBC (even particle numbers) a significant overlap with the $M$ and $M+1$  rotating states is only obtained for $n=\lbrace M-1,M,M+1,M+2\rbrace$ states, with $c_{M(M-1)}^{\vphantom{-}} c_{(M+1)(M-1)}^* \approx - c_{M(M+2)}^{\vphantom{-}} c_{(M+1)(M+2)}^*$  and  $c_{MM}^{\vphantom{-}} c_{(M+1)M}^* \approx - c_{M(M+1)}^{\vphantom{-}} c_{(M+1)(M+1)}^*$. There will therefore be a significant contribution to the time-dependent oscillation for $N=M$. For  $N=M+2$ and higher particle numbers the time-dependent oscillation becomes small again as the contributions from the $n=\lbrace M-1,M\rbrace$ terms are canceled out by the contributions from the $n=\lbrace M+1,M+2\rbrace$ terms. For PBC (odd particle numbers) only the states $n= \lbrace M-2,M-1,M+2,M+3 \rbrace$ have significant overlap with the the $M$ and $M+1$  rotating states, with $c_{M(M-2)}^{\vphantom{-}} c_{(M+1)(M-2)}^* \approx - c_{M(M+2)}^{\vphantom{-}} c_{(M+1)(M+2)}^*$  and  $c_{M(M-1)}^{\vphantom{-}} c_{(M+1)(M-1)}^* \approx - c_{M(M+3)}^{\vphantom{-}} c_{(M+1)(M+3)}^*$. This means that there will be a significant contribution for $N=M-1$ and $N=M+1$. For $N=M+3$ and higher particle numbers the time-dependent oscillation becomes small again as the contributions from $n= \lbrace M-2,M-1\rbrace$ and $n= \lbrace M+2,M+3 \rbrace$ cancel out. The analysis for states close to $2M$ is exactly the same, that is only $ F_{2M(2M+1)}(t) = F_{(2M+1)2M}(t) \neq 0$ are important and they contribute in the same pattern. However, in this case a deeper lattice ($V_0 \approx 0.8 E_r$) is required for significant gaps to be introduced. The large finite values of the average momentum for commensurate particle numbers visible in Fig.~\ref{fig:currentasfunctionofspeed} are therefore entirely due to a time-dependent oscillation induced when quenching to the rotating lattice system. The time-dependence of $ F_{(M)(M+1)}(t)$ is periodic with $t_0$, while $F_{(2M)(2M+1)}(t)$ is periodic with $\frac{1}{2}t_0$. The latter periodicity is because the relevant SP eigenfunctions from which $F_{(2M)(2M+1)}(t)$ is obtained are also periodic with $\psi_{j\geq 2M,\Omega}(x-vt) \approx \psi_{j \geq 2M,\Omega}(x-v(t+\frac{1}{2}t_0))$. In these cases the periodicity associated with the gap energy, $T=\frac{2 \pi \hbar}{E_{M+1}-E_{M}},\frac{2 \pi \hbar}{E_{2M+1}-E_{2M}}$ is very large, as the gap is very small (see Fig.~\ref{fig:oddevenspectrum}). This long time scale oscillation is therefore obscured by the more important short time scale oscillation, see Fig.~\ref{fig:instantaneousplotsshallowlattice}.

For slightly deeper lattices ($0.02 E_r <V_0 < 1.5 E_r$), the $d>0$ contributions start to become important and multiple frequency contributions are the main source of the finite momentum we see close to commensurability in Fig.~\ref{fig:dynamicphasediagram}. The difference between the odd and even particle number spectra close to $\Omega=2 \eta+1$ and  $\Omega=2 \eta$ also becomes smaller as the gap between the states $M-d$ and $M+1+d$ increases and the energy states within each band move closer to each other. The odd-even effect therefore eventually disappears.

\subsubsection{Intermediate and deep lattices}
For a lattice of intermediate depth $1.5 E_r < V_0 <10 E_r$ the finite momentum observed close to commensurability in Fig.~\ref{fig:dynamicphasediagram} stems mostly from the time-independent part of the average momentum. In these intermediate lattices the terms in the sum $\sum_{n}\left(\sum_{j} |c_{jn}|^2 \mathrm{Im} [F_{jj}]\right)$ are finite, but negative for small values of $n<M$ and positive as $n$ gets close to $M$, which means that the resulting sum is close to zero, leading to $K(t)=\frac{N \hbar q}{m L}+K_t(t)$, only when $N \approx M$. As we go towards the deep lattice limit ($V_0 > 10 E_r$) the states in each each band become almost degenerate which results in $F_{jj} \approx 0$ leading to $K(t)=\frac{N \hbar q}{m L}+K_t(t)$, which explains the linear increase with $q$, even for incommensurate particle numbers, observed in Fig.~\ref{fig:currentasfunctionofspeed} for $V_0= 20 E_r$. For deep lattices, $F_{(M-d)(M+1+d)}(t)$ is quite large for all possible values of $d$ and despite the small overlaps $c_{(j>M)(n<M)}$ the sum of these contributions gives rise to quite large oscillations. All of them have a similar periodicity corresponding to the band gap energy $T_{M(M+1)}=\frac{2 \pi \hbar}{E_{M+1}-E_{M}}$. As the gap energy is large, these are rapid oscillations on short timescales and they therefore dominate compared to the $t_0$ periodicity for $F_{(M-d)(M+1+d)}(t)$, although this periodicity can still be observed in the instantaneous average momentum. The oscillations correspond to the excitation of particles to higher energy-bands within each site, due to the rotation of the lattice. So although the particles are transported along with the lattice on average, these on-site excitations leads to time-dependent dynamics within each site, as these have a finite width due to the continuum nature of the lattice. This type of oscillation is observed in Fig.~\ref{fig:instantaneousplotsdeeplattice}.

\FloatBarrier

\end{document}